\documentclass{webofc}
\usepackage[varg]{txfonts}   
\usepackage{tikz}
\usepackage{mathrsfs}
\usepackage{feynmp}
\usepackage{tensor}
\usepackage{amsmath}
\usepackage{wasysym}
\usepackage{amsfonts}

\usepackage{yfonts}

\usepackage[colorlinks=true,linkcolor=black,citecolor=black,urlcolor=blue,filecolor=black]{hyperref}

\usepackage{color}
\definecolor{rougef}{rgb}{0.56,0,0}
\definecolor{vertf}{rgb}{0,0.5,0}
\definecolor{bleuf}{rgb}{0,0,0.8}
\definecolor{violetf}{rgb}{0.5,0,0.5}

\def\pe{\prime}
\def\3s{{s \choose 3}}
\def\4s{{s \choose 4}}
\def\5s{{s \choose 5}}
\def\6s{{s \choose 6}}

\def\12{\dfrac{1}{2}}

\def\ft{\footnote}

\def\2{\ell_2}

\def\pr{\partial}

\def\prd{\partial \cdot}

\def\scri{\mathscr{I}}

\def\scrip{\mathscr{I}^+}

\def\be{\begin{equation}}
\def\ee{\end{equation}}
\def\bea{\begin{eqnarray}}
\def\eea{\end{eqnarray}}
\def\ba{\begin{array}}
	\def\ea{\end{array}}
\def\bec{\begin{center}}
	\def\ec{\end{center}}

\newcommand{\bin}[2]{{#1 \choose #2}}

\def\a{\alpha} 
\def\b{\beta}  
\def\g{\gamma} 

\def\d{\delta} 

\def\e{\epsilon}

\def\m{\mu}

\def\r{\rho}

\def\vf{\varphi}

\def\cM{{\cal M}}

\def\cO{{\cal O}}

\begin{document}
\begin{fmffile}{diagram}
\title{Asymptotic symmetries and charges at null infinity: from low to high spins}
%
%

\author{\firstname{Andrea} \lastname{Campoleoni}\inst{1}\fnsep\thanks{\email{campoleoni@itp.phys.ethz.ch}} \and
        \firstname{Dario} \lastname{Francia}\inst{2,3}\fnsep\thanks{\email{dario.francia@roma3.infn.it}} \and
        \firstname{Carlo} \lastname{Heissenberg}\inst{4}\fnsep\thanks{\email{carlo.heissenberg@sns.it}}
}

\institute{Institut f\"ur Theoretische Physik, ETH Z\"urich,
	Wolfgang-Pauli-Strasse 27, 8093 Z\"urich, Switzerland
\and
		   Museo Storico della Fisica e Centro Studi e Ricerche E. Fermi,
		   Piazza del Viminale 1, I-00184 Roma, Italy
\and
           Roma Tre University and INFN, Via della Vasca Navale 84, I-00146 Roma, Italy 
\and
           Scuola Normale Superiore and INFN, Piazza dei Cavalieri 7, I-56126 Pisa, Italy
          }

\abstract{%
  Weinberg's celebrated factorisation theorem holds for soft quanta of arbitrary integer spin. The same result, for spin one and two, has been rederived assuming that the infinite-dimensional asymptotic symmetry group of Maxwell's equations and of asymptotically flat spaces leave the S-matrix invariant. For higher spins, on the other hand, no such infinite-dimensional asymptotic symmetries were known and, correspondingly, no a priori derivation of Weinberg's theorem could be conjectured. In this contribution we review the identification of higher-spin supertranslations and superrotations in $D=4$ as well as their connection to Weinberg's result. While the procedure we follow can be shown to be consistent in any $D$,  no infinite-dimensional enhancement of the asymptotic symmetry group emerges from it in $D>4$, thus leaving a number of questions unanswered.
}

\maketitle

\noindent
{\it Based on the talk given by D.F. at Quarks-2018 -- XXth International Seminar on High-Energy Physics. Valday, Russia. May 27 -- June 2, 2018.}


\section{Introduction}
\label{intro}
In this contribution we review the study of large higher-spin gauge transformations on Minkowski backgrounds, of the corresponding conserved charges, and of their relations to soft theorems carried out in \cite{super,noi_charges}. 
The main highlights of our analysis are the following: 
\begin{itemize}
\item Upon imposing suitable falloff conditions at null infinity, providing a finite, non-vanishing asymptotic energy flux, the residual gauge transformations of Fronsdal fields \cite{fronsdal} on a four-dimensional Minkowski background generate an infinite-dimensional symmetry algebra, comprising higher-spin counterparts of supertranslations and superrotations (the latter explicitly computed only for $s=3$). The higher-spin supertranslation Ward identities, in particular,  can be shown to imply Weinberg's soft theorem to leading order \cite{Weinberg_64, Weinberg_65}, in strict analogy with recent results concerning  spin one and spin two gauge fields \cite{Strominger_Invariance, Strominger_Weinberg}.
\item Extending the analysis  to arbitrary values of the space-time dimension, we provided falloff conditions in a Bondi-like gauge  and tested their consistency against the computation of the corresponding conserved charges, that result to be finite and non vanishing. The asymptotic symmetry algebra in $D>4$, on the other hand,  appears to only comprise the global solution to the Killing tensor equations, while not displaying any infinite-dimensional enhancement.
\end{itemize}

The key observation that triggered our work was that Weinberg's results of \cite{Weinberg_64, Weinberg_65}, relating amplitudes involving the emission or absorption of particles in the limit of vanishing momentum, hold for soft massless particles of any spin and possibly in any space-time dimension $D$. It was then natural to ask whether or not also for $s \geq 3$ one could connect those results to the Ward identities of some yet-to-be identified infinite-dimensional symmetry, in the spirit of the recent findings  on the role played by supertranslations \cite{BMS, Sachs_Waves, Sachs_Symmetries} (see also \cite{Geroch_Lectures, Ashtekar_Lectures}) and $U(1)$ large gauge transformations in amplitudes involving soft gravitons and soft photons \cite{Strominger_Invariance, Strominger_Weinberg, Strominger_YM, Strominger_QED, Campiglia_QED, Avery-Schwab, Barnich_BMS/CFT, Barnich_Revisited, Barnich_Charge, Hamada:2018vrw}. (See also \cite{Strominger_rev} and references therein.)

In order to investigate this issue we assigned a set of Bondi-like falloff conditions to the spin-$s$ potentials, to be interpreted as a combination of local gauge fixing and consequences of the equations of motion, and we computed the gauge symmetries that keep  those conditions. Among them, in $D=4$ we identified an infinite-dimensional class of large gauge symmetries of the Fronsdal action which provide proper counterparts of spin-$2$ supertranslations, and showed how the corresponding Ward identities allow indeed to derive Weinberg's soft theorem for arbitrary integer spin. Higher-spin supertranslations, however, provide only a special class of asymptotic symmetries for our systems. For $s=3$, in addition, we also exhibited the general transformations preserving our Bondi-like falloff conditions and showed the existence of additional infinite families of asymptotic symmetries, naturally identified with higher-spin generalisations of superrotations \cite{Barnich_Revisited,Barnich_BMS/CFT}.  The higher-spin four-dimensional analysis was presented in \cite{super} and is reviewed in Section~\ref{sec: as-st}.

Section \ref{sec: Charges} is instead devoted to illustrating the results of \cite{noi_charges}, mainly concerned with the arbitrary-dimensional case. First, we provide a full analysis of the non-linear Yang-Mills theory in any dimension where, with respect to the existing literature \cite{Einstein-YM Barnich}, we add the explicit computation of the charges\ft{Although we do not include this aspect in the present review, let us mention that for the three-dimensional case, already discussed in \cite{Maxwell d=3 Barnich}, in \cite{noi_charges} we also included the contribution of radiation.} (see also \cite{Strominger_YM,Strominger_QED,Campiglia_QED,Mao_em,Mao_note,Strominger "Kac",Strominger Color,Adamo Casali} for the four-dimensional case).  For the general case of spin $s$ we assume that the asymptotic behaviour of the field components can be parameterised by an expansion in  powers of the radial coordinate\ft{For the case of Yang-Mills theory in $D=3$ logarithmic dependence was also taken into account. A covariant alternative to the use of radial falloffs in Bondi coordinates would entail exploiting  the notion of conformal null infinity. The construction of the latter, however, was shown to be obstructed in odd-dimensional spacetimes containing radiation, because of singularities appearing in the components of the Weyl tensor of the unphysical space \cite{hollands2004}. Differently, our computation of the falloffs, although non covariant,  results in an exploration of null infinity that can be implemented in all dimensions \cite{gravity_anyD}.}, complemented with the assumption of Bondi-like gauge conditions. We take into account that, whenever $D$ is odd and greater than four, in order for both the radiation and the Coulomb parts of the solution to be accounted for, one finds that the expansion in powers of $r$ requires both integer and half-integer exponents to be considered. Differently, only integer powers of $r$ are needed whenever $D$ is even\ft{Radiation and Coulombic contributions behave as $r^{(2-D)/2}$ and $r^{3-D}$, respectively, and thus coincide in $D=4$.}. 

 Leading and subleading falloffs are determined by solving the equations of motion, while their consistency relies on checking that the energy flowing to null infinity per unit of retarded time is indeed finite. Once the falloffs are determined, we compute the asymptotic symmetries and the corresponding charges, while also verifying finiteness of the latter. The resulting symmetry algebra in $D > 4$, however, turns out to consist just of the global solutions to the Killing tensor equations and no infinite-dimensional enhancement emerges, thus posing a number of questions both on the ultimate physical meaning of our findings and on the generality of the whole procedure. On the one hand, our result is in agreement with similar conclusions drawn for spin two in previous works \cite{angular-momentum,gravity_evenD_2}. On the other hand, it is manifestly at odds with the validity of Weinberg's soft theorems in any $D$, unless one is willing to accept that the same class of physical phenomena admit a universal interpretation in terms of symmetries just in one specific dimension, while still holding in infinitely many other dimensionalities where the same symmetries simply do not exist.

In order to address this issue,  in \cite{Pate_Memory}  an alternative treatment of boundary conditions for spin two was proposed, together with a different gauge, allowing for infinite-dimensional symmetries for linearised gravity to be identified in any even dimension in subleading contributions in $1/r$. These symmetries, in their turn, were shown to be responsible for both Weinberg's theorem in $D = 2k$ and for even-dimensional counterparts of the memory effect. (See also \cite{Kapec:2015vwa,Garfinkle:2017fre,Mao_evenD,Campiglia_scalars} for earlier discussions on the matter.) It is therefore possible to envision that a similar procedure could be carried out for spin one and for higher spins, in order to address the corresponding questions about the ultimate origin of Weinberg's soft theorem in $D>4$ and  of electromagnetic, Yang-Mills or, perhaps, higher-spin memory \cite{ACDmemory}.

Among the outstanding issues, it ought to be stressed that our linearised analysis does not allow one to get a concrete grasp on the properties of the putative non-Abelian algebra underlying our findings for higher spins. This aspect should be relevant in particular in order to assess their role in the high-energy regime of string scattering amplitudes, where one may expect to see remnants of such symmetries if string theory can really be viewed as a broken phase of some higher-spin gauge theory (see \emph{e.g.}~\cite{Sagnotti_review} and references therein). Once again, the investigation on the possible infinite-dimensional enhancement of global asymptotic symmetries for all spins in $D>4$ would be of special relevance in this respect.


\section{Higher-spin asymptotic symmetries and soft theorems}
\label{sec: as-st}

 In \cite{Weinberg_64, Weinberg_65}, Weinberg considered  the $S$-matrix element $S_{\beta\alpha} (\mathbf q)$, involving arbitrary asymptotic particle states $\a$ and $\b$ together with an extra soft massless particle of $4-$momentum \mbox{$q^{\, \mu}=  (\omega, \mathbf q) \to 0$} and helicity $s$. The two main contributions to this process are schematically encoded in  the following picture:
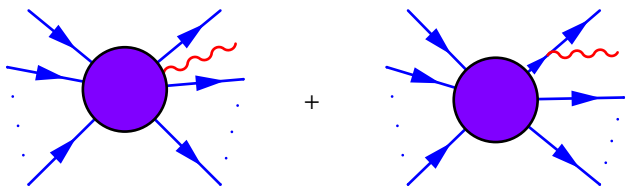
\begin{align*}
\begin{gathered}
\begin{fmfgraph*}(90, 65)
\fmfleft{i1,d1,i2,i3}
\fmfright{o1,d2,d3,o2,o4,o3}
\fmf{fermion, fore=blue}{i1,v1}
\fmf{fermion, fore=blue}{i2,v1}
\fmf{fermion, fore=blue}{i3,v1}
\fmf{fermion, fore=blue}{v1,o1}
\fmf{fermion, fore=blue}{v1,o2}
\fmf{fermion, fore=blue}{v1,o3}
\fmf{dots, fore=blue}{i2,i1}
\fmf{dots, fore=blue}{o1,o2}
\fmfset{dot_len}{4.7mm}
\fmf{photon, fore=red, tension=0}{v1,o4}
\fmfv{d.sh=circle,d.f=empty,d.si=.35w,b=(.5,,0,,1)}{v1}
\end{fmfgraph*}
\end{gathered}
\qquad + \qquad 
\begin{gathered}
\begin{fmfgraph*}(90, 65)
\fmfleft{i1,d1,i2,i3}
\fmfright{o1,d2,o2,op,o3}
\fmf{fermion, fore=blue}{i1,v1}
\fmf{fermion, fore=blue}{i2,v1}
\fmf{fermion, fore=blue}{i3,v1}
\fmf{fermion, fore=blue}{v1,o1}
\fmf{fermion, fore=blue}{v1,o2}
\fmf{fermion, fore=blue}{v1,o}
\fmf{fermion, fore=blue}{o,o3}
\fmf{photon, fore=red, tension=0}{o,op}
\fmfv{d.sh=circle,d.f=empty,d.si=.35w,b=(.5,,0,,1)}{v1}
\fmf{dots, fore=blue}{i2,i1}
\fmf{dots, fore=blue}{o1,o2}
\fmfset{dot_len}{3.95mm}
\end{fmfgraph*}
\end{gathered}
\end{align*}
where the second diagram, in particular, provides the leading contribution to the process. Using only the Lorentz invariance and the pole structure of the $S$ matrix, he showed that the latter contribution takes a factorised form that, in the notation of \cite{Weinberg_64, Weinberg_65}, can be written as
\be \label{wfact}
\lim_{\omega\to0^+}\omega\, S_{\beta \alpha}^{\pm s}(\mathbf q) = 
-\lim_{\omega\to0^+}\left[\, \omega\sum_i \eta_i^{\phantom{(s)}}\!\!\! g_i^{(s)} \frac{(p_i \cdot \varepsilon^{\pm}(\mathbf q))^s}{p_i\cdot q}\,\right]S_{\beta\alpha} \,,
\ee
with $\eta_i$ being $+1$ or $-1$ according to whether the particle $i$ is incoming or outgoing, while $S_{\alpha\beta}$ denotes the $S$ matrix for the same process without the soft particle.

Recasting Weinberg's soft theorem from the momentum-space form \eqref{wfact} to its position-space counterpart in retarded Bondi coordinates\ft{The Minkowski metric in such coordinates reads 
\[
ds^2 = - du^2 - 2 du dr + 2 r^2\, \gamma_{z\bar{z}} dz d\bar{z} \,, \qquad\quad \g_{z\bar{z}} = 2\left(1+z\bar{z}\right)^{-2} \,.
\]
} $u$, $r$, $z$ and $\bar z$  one has
\be\label{WeinbergPSPACE}
\lim_{\omega\to0^+}\omega\, S_{\beta\alpha}^{+s}  = (-1)^{s}\,{2^{\frac{s}{2}-1}}
(1+z\bar z)\left[\, \sum_i \eta_i^{\phantom{(s)}}\!\!\! g_i^{(s)} 
\frac{(E_{i})^{s-1}(\bar z-\bar{z}_i)^{s-1}}{(z-z_i)(1+z_i\bar z_i)^{s-1}}
\,\right]S_{\beta\alpha} \, ,
\ee
where $E_i$ and $(z_i,\bar z_i)$ characterise the massless particles scattered to null infinity\footnote{The coordinates $(z_i,\bar z_i)$ identify the point on the celestial sphere where a wave packet of spatial momentum centred around $p_i$ becomes localised at late times, while $E_i$ denotes the corresponding energy.}. One would like to understand the origin of \eqref{WeinbergPSPACE} as due to some symmetry principle for all spins, and not just for spin one and spin two.

\subsection{Higher-spin supertranslations and Weinberg's theorem}\label{subsec: SSTransl}

The Fronsdal action \cite{fronsdal} is invariant under the transformation of the gauge potential
\be
\delta \varphi_{\mu_1 \ldots \mu_s} 
= \partial_{(\mu_1} \e_{\mu_2\ldots \mu_s)}\, ,
\ee
with a doubly traceless field and a traceless gauge parameter. Our Bondi-like gauge is summarised by the conditions
\begin{align}
&\varphi_{r \mu_2 \ldots \mu_s}=0=\varphi_{z \bar z \mu_3 \ldots \mu_s}  \label{constr_spin-s} \, ,\\[5pt]
&\varphi_{uu\ldots u \underbrace{\scriptstyle zz \ldots z}_d} = r^{d-1} B_{zz\ldots z}(u,z,\bar z) + \cO(r^{d-2})  \label{boundary_spin-s}\, ,
\end{align}
for $d=0,\ldots,s$, together with their conjugates. These ensure in particular that the field be traceless, $\vf^{\, \pe}_{\, \m_3 \, \ldots \m_s} = 0$, so that the Fronsdal equations take the Maxwell-like form \cite{lowspin, ML}
\be
\Box \,\varphi_{\mu_1 \ldots \mu_s} \, - \, \pr_{(\m_1} \, \prd \varphi_{\mu_2 \ldots \mu_s)} \, = \, 0 \, .
\ee
We shall denote with $\varphi^p_{d ,c}$ and $\e^p_{d,c}$ the field  and the gauge parameter components, respectively, where $p$ is the number of ``$u$'' indices, $d$ is the number of ``$z$'' indices appearing without $\bar z$ counterpart and $c$ is the number of pairs ``$z \bar z$''. 

To begin with, we look for the residual gauge freedom which preserves \eqref{constr_spin-s} and \eqref{boundary_spin-s} with $u-$independent traceless parameters displaying power-like dependence on $r$. As it turns out, it admits the following parametrisation: 
\begin{align}\label{pd0}
\e^p_{d,0}=&- \frac{r^d D^d_zT_p(z, \bar z)}{\prod_{k=1}^d (s-p-k)}\,,\\
\label{c+1c}
\e^p_{d,c+1} =& - \frac{r^2}{2}\,\gamma_{z\bar z} \left( \e^p_{d,c} - 2\, \e^{p+1}_{d,c}\right),
\end{align}
where $D_z$ denotes  the covariant derivative given by the metric $\g_{z \bar{z}}$ on the Euclidean unit two-sphere, while $T_p(z, \bar z)$ for $p=0,\ldots,s-1$ is a set of angular functions satisfying
\be
\label{p+1p}
T_{p+1} =\frac{s-p}{s[s-(p+1)]}\,T_p + \frac{1}{[s-(p+1)]^2}\,D^zD_zT_p\,.
\ee
Therefore, this family of residual gauge transformations is defined recursively in terms of a single angular function $T_0(z, \bar z) \equiv T(z,\bar z)$, in close analogy with the cases of spin one and of spin-two supertranslations. 

Assuming that a suitable diagonal subgroup of the product of these infinite-dimensional symmetries at $\scri^-$  and $\scri^+$ be a group of symmetries of the $S$ matrix, we can write the corresponding Ward identity schematically as follows
\be \label{ward}
[Q_s + Q_h, S] = 0\, ,
\ee
where $Q_s + Q_h$ is the total charge associated to the previous asymptotic symmetry. We would now like to show that \eqref{ward} implies \eqref{WeinbergPSPACE}.
Using the auxiliary boundary condition $(D^z)^sB_{z\ldots zz} = (D^{\bar z})^sB_{\bar z\ldots\bar z \bar z}$ and integrating by parts, the charge corresponding to our family of large gauge transformations can be written at $\scri^+$ as 
\be \label{Q-total}
Q^+ = \underbrace{(-1)^s\frac{s}{2(s-1)!}\int_{\scrip} \partial_{\bar z} T (D^z)^{s-1}\partial_u B_{z\ldots zz} d^2z du}_{Q^+_s} - \underbrace{\frac{s}{2} \int_{\scrip} \gamma_{z\bar z} J(u, z, \bar z)d^2z du}_{Q^+_h}  \,.
\ee
One can then show that, assuming the action of $Q_h$ on scalar matter fields be given by
\be
[Q^+_h, \Phi] = \frac{s}{2}\,g_i^{(s)}T(i\partial_u)^{s-1}\Phi\, ,
\ee
and by judiciously choosing 
\be
T(z, \bar z) = \frac{1}{w-z}\left(\frac{1+w\bar z}{1+z\bar z}\right)^{\!s-1}\, ,
\ee
the large$-r$ behaviour of \eqref{ward} yields indeed
\be
\lim_{\omega\to0^+} \left[\,\omega\,\langle \text{out} | a^{\text{out}}_+S | \text{in}\rangle \,\right] = 
(-1)^s 2^{s/2-1} (1+z\bar z) \sum_i \eta_i\, \frac{g_i^{(s)}E_n^{s-1}}{z-z_i} \left(\frac{\bar z - \bar z_i}{1+z_i \bar z_i}\right)^{\!s-1} .
\ee
Thus, Weinberg's factorisation can be understood as a manifestation of an underlying spin-$s$ large gauge symmetry acting on the null boundary of Minkowski spacetime. For more details on these computations see also \cite{carlo_tesi}. 

\subsection{Higher-spin superrotations}\label{sec:superrotations}

Looking for the most general asymptotic symmetry\footnote{For $D > 3$, in \cite{super,noi_charges} we actually studied only symmetries with parameters admitting an expansion in powers of $r$ and $u$, excluding for simplicity possible logarithmic dependences even if they can appear in gravity (see \emph{e.g.}~\cite{Barnich_BMS/CFT}).} still preserving the Bondi-like gauge defined by equations \eqref{constr_spin-s} and \eqref{boundary_spin-s} and restricting to the case of spin three for the sake of simplicity, one finds that the corresponding gauge parameter can be expressed in terms of the following quantities defined on the celestial sphere: $T(z,\bar z)$, $K_{zz}(z, \bar z)$, $K_{\bar z \bar z}(z, \bar z)$, $\rho_z(z,\bar z)$, $\rho_{\bar z}(z, \bar z)$ (we refer the reader to \cite[Section 6 and Appendix B]{super} for further details). 

The function $T(z, \bar z)$ is not constrained at all and this leads to the higher-spin supertranslations discussed in the previous subsection. The tensors $\r$ and $K$ are instead bounded to satisfy appropriate differential equations. Remarkably, when the dimension of the spacetime is four, locally they both admit infinitely many solutions. The equation for $K_{zz}$, $K_{\bar z \bar z}$ is indeed the rank-2 conformal Killing equation \cite{conformal-killing}. Being traceless, it only admits two non-trivial components that, using a holomorphic parameterisation of the metric, read 
\be
\pr_{\bar{z}} K^{zz} = 0 \, , \qquad 
\pr_{z} K^{\bar{z}\bar{z}} = 0 \, .
\ee
Its solutions are therefore locally characterised by a holomorphic and an antiholomorphic function
\be \label{sol2-K}
K^{zz} = K(z) \, , \qquad 
K^{\bar{z}\bar{z}} = \tilde{K}(\bar{z}) \, , \qquad
K^{z\bar{z}} = 0 \, .
\ee
In a similar fashion, the equations for $\rho_z$, $\rho_{\bar z}$,
\be
D_z D_z \rho_{z}=0\,,\qquad D_{\bar z} D_{\bar z} \rho_{\bar z}=0\, ,
\ee
can be cast in the form
\be
\pr_{\bar{z}}\! \left( \g^{z\bar{z}} \pr_{\bar{z}} \r^z \right) = 0 \, , \qquad
\pr_{z}\! \left( \g^{z\bar{z}} \pr_{z} \r^{\bar{z}} \right) = 0 \, ,
\ee
and are solved by
\be \label{sol2-r}
\r^z = \a(z)\, \pr_z k(z,\bar{z}) + \b(z) \, , \qquad
\r^{\bar{z}} = \tilde{\a}(\bar{z})\, \pr_{\bar{z}} k(z,\bar{z}) + \tilde{\b}(\bar{z}) \, ,
\ee
where $k(z,\bar{z})$ is the K\"ahler potential for the $2$-dimensional metric $\gamma_{z \bar z}$ on the unit sphere\ft{For instance, one can choose $k(z,\bar{z}) = 2\log(1+z\bar{z})$.}, while $\a(z)$ and $\b(z)$ are instead arbitrary holomorphic functions. Similar considerations apply to the antiholomorphic sector.

 In the case of the  BMS algebra, supertranslations and superrotations  can be considered as infinite-dimensional enhancements of the Killing symmetries associated to the Poincar\'e generators $P^\mu$ and  $M^{\mu\nu}$. For higher spins the global solutions of the Killing tensor equation (imposing $\d \vf_{\m_1 \ldots \m_s} = 0$ rather than the preservation of our falloffs) are in one-to-one correspondence with the traceless projections of the combinations $P^{(\mu}P^{\nu)}$, $P^{(\mu}M^{\nu)\rho}$ and $M^{\rho(\mu}M^{\nu)\sigma}$, candidate spin-three generators of a putative higher-spin algebra possessing a  Poincar\'e subalgebra (see {\it e.g.}\ \cite{flat-algebras_1,flat-algebras_2} for  discussions on higher-spin algebras possibly related to  Minkowski space). The asymptotic symmetries generated by $T$, $\rho_z$, $\rho_{\bar z}$, $K_{zz}$, $K_{\bar z \bar z}$ could thus be interpreted as their corresponding infinite-dimensional enhancements. 


\section{Asymptotic symmetries and charges in any dimension}
\label{sec: Charges}

We now move to the case of spacetimes of arbitrary dimensions, adopting the following notation for the retarded Bondi coordinates:   $(x^\mu)=(u,r,x^i)$, where $x^i$, for $i=1,2,\ldots,n$, denotes the $n := D-2$ angular coordinates on the sphere at null infinity\footnote{In this contribution we focus on $D > 3$, \emph{i.e.}, on $n > 1$. We refer to \cite{noi_charges} for the analysis of the three-dimensional ($n=1$) case.}. In these coordinates the Minkowski metric reads 
\be \label{bondi-coord}
ds^2 = - du^2 - 2 du dr + r^2 \gamma_{ij}\, dx^i dx^j\,,
\ee
where $\gamma_{ij}$ is the metric of the Euclidean $n$-sphere. We first review the computation of asymptotic symmetries and charges in the non-linear Yang-Mills theory to then extend the analysis to spin-$s$ free fields.

\subsection{Spin one}

We shall denote the Yang-Mills field by
$
\mathcal A_\mu := \mathcal A_\mu^A T^A\, ,
$ 
where the $T^A$ are the generators of a compact Lie algebra $\mathfrak g$, whose gauge transformation is  $ \delta_\epsilon\mathcal A_\mu = \nabla_{\!\mu} \epsilon + [\, \mathcal A_\mu, \epsilon \,] $.  
As in \eqref{constr_spin-s}, we enforce the radial gauge
\be \label{bondi1}
\mathcal A_r=0 \,,
\ee
which completely fixes the gauge in the bulk.

For $D > 3$ we consider field configurations  $\mathcal A_\mu$ whose asymptotic null behaviour is captured by an expansion in powers of $1/r$.
By analysing the leading-order equations of motion, two possible types (or ``branches'') of solutions arise retaining relevant physical meaning. The first one corresponds to \emph{radiation}, with the familiar leading falloffs 
\be
\mathcal A_u = A^{(n/2)}_u (u, x^k)\, r^{-n/2} \, , \qquad
\mathcal A_i = A^{(n/2-1)}_i (u, x^k)\, r^{-n/2+1} 
\ee
of a spherical wave,  giving rise to a finite flux of energy at null infinity per unit of retarded time. The latter, on the other hand, leads to \emph{Coulomb-type} solutions with the characteristic leading falloffs
\be
\mathcal A_u =  A^{(n-1)}_{u}(u, x^i)\, r^{1-n} \, , \qquad
\mathcal A_u =  \cO(r^{1-n})
\ee
of the Coulomb potential, providing a finite contribution to the colour charge.

Let us stress that the presence of two distinct branches of  solutions, radiation and Coulombic, is apparent only for $D>4$, while in in $D = 4$ they effectively coincide. Furthermore, since the exponent appearing in $r^{-n/2}$ is non-integer for odd dimensions, one needs to consider two separate $1/r$ expansions in the case of odd-dimensional spacetimes in order to account for both Coulombic and radiation terms.

The ultimate goal of the analysis is the study of the compatibility between the falloff conditions and the on-shell finiteness of two relevant physical quantities: the charge associated to asymptotic symmetries, and the energy flux at infinity.

The colour charge, in particular, receives contributions that diverge off-shell, such as, for instance, a term behaving like $r^{n/2-1}$ as $r\to\infty$ from the leading radiation term $A_u^{(n/2)} r^{-n/2}$. Upon performing a detailed analysis of the equations of motion, however, one can prove that all the potentially divergent terms actually cancel on shell, thus ensuring the finiteness of the colour charge to all orders. This is eventually expressed as the following integral over the $n$-sphere at each point $u$,
\be\label{ColoreD>4}
Q^A(u)=(n-1)\int_{S_u} \mathrm{tr} \left( A_u^{(n-1)} T^A \right)\, d\Omega_n\,,
\ee
which only involves the Coulombic order.
The energy flux, on the other hand, is expressed as
\be \label{power}
\mathcal P(u)=-\int_{S_u} \gamma^{ij}\, \mathrm tr\left(\partial_u A_i^{(n/2-1)} \partial_u A_j^{(n/2-1)}\right)d\Omega_n\,
\ee
and consistently only involves the leading radiation terms. The nonlinearities  induce however an interplay between radiation and Coulombic terms: namely, radiation leads to a leakage of colour charge across null infinity expressed by 
\be
\frac{d}{du}\, Q^A(u) = \int_{S_u} \gamma^{ij} \left[A_i^{(n/2-1)}, \partial_u A_j^{(n/2-1)}\right]^A d\Omega_n\,.
\ee
Let us observe that the above colour charges  form a representation of the underlying algebra: for any $D\geq 2$, since $\delta_\epsilon A_u= [A_u,\epsilon]$,
\be\label{commutation_rel}
[Q_{\epsilon_1},Q_{\epsilon_2}]=\delta_{\epsilon_1} Q_{\epsilon_2}= \int_{S_u} \mathrm{tr}\left([A_u, \epsilon_1]\epsilon_2\right) d\Omega_n =  \int_{S_u} \mathrm{tr}\left(A_u [\epsilon_1,\epsilon_2]\right) d\Omega_n= Q_{[\epsilon_1,\epsilon_2]}\, .
\ee
While \eqref{commutation_rel} holds in any dimension, it should be stressed that, when $D>4$, the corresponding charge algebra coincides with $\mathfrak g$, given that only constant gauge parameters preserve our falloffs, whereas in $D=3, 4$ it is in fact an infinite-dimensional Kac-Moody algebra, owing to the arbitrary parameters $\epsilon(x^1,x^2)$ and $\epsilon(u,\phi)$ preserving the falloffs. In particular, we note the absence of a central charge, which could however emerge by performing the analysis for the linearised theory around a nontrivial background, as pointed out in \cite{Avery-Schwab}.

\subsection{Arbitrary spin}

Although technically more involved, completely analogous results can be derived for the case of a linearised field carrying an arbitrary integer spin $s$, with the exception of the charge flux across null infinity, which would require control of the nonlinear theory.

To streamline the presentation, from now on groups of symmetrised indices will be substituted by a single letter with a label denoting the total number of indices. For instance, $\vf_{r\ldots r\, i_1\ldots i_{k}} \to \vf_{r_{s-k} i_{k}}$. The spin-$s$ analogues of the radial gauge conditions \eqref{constr_spin-s} and \eqref{bondi1} then read
\be
\vf_{r\,\mu_{s-1}} = 0 \, , \qquad \g^{ij} \vf_{ij\,\m_{s-2}} = 0 \, .
\ee
Locally these constraints can be imposed with an on-shell gauge fixing; let us stress, however, that for spin two this choice is likely to be at the origin of the discrepancy between our findings and those of \cite{Pate_Memory}.

Assuming an expansion in powers of $1/r$, the equations of motion lead to two branches of solutions also in the general, spin-$s$ case.
The first one corresponds to \emph{radiation}: it admits the leading falloffs 
\be \label{boundary-cond-s}
\vf_{u_{s-k}i_k} = r^{-\frac{n}{2}+k}\, a_{s}[k]\, (D\cdot)^{s-k} C_{i_k}{}^{\!\!\!(s)}(u,x^j) \, ,
\ee
where the $a_{s}[k]$ are some coefficients that have been determined in \cite{noi_charges}.  At the leading order, this branch thus only depends on the traceless tensor $C^{(s)}_{i_s}(u,x^j)$ and its divergences, and it gives a finite energy flux per unit retarded time at infinity:
\be
\mathcal P(u)=\int_{S_u} \gamma^{i_1 j_1} \ldots \gamma^{i_s j_s}\, \partial_u C^{(s)}_{i_1 \ldots i_s} \partial_u C^{(s)}_{j_1 \ldots j_s} d\Omega_n\,.
\ee 
The second branch corresponds to \emph{Coulombic} solutions with leading falloffs of the form
\be \label{coulomb-s}
\vf_{u_{s-k}i_k} = r^{1-n} \sum_{l\,=\,0}^k b_{s}[k,l]\, u^l \underbrace{D_{i} \cdots D_i}_{\textrm{$l$ terms}} \cM_{i_{k-l}}{}^{\!\!\!\!\!\!\!(k-l)}(x^j) + \cdots \, , 
\ee
where we omitted the terms enforcing the traceless projection. It thus depends on $k$ traceless tensors $\cM_{i_{k}}{}^{\!\!\!(k)}(x^j)$ of ``integration constants''  and gives a finite contribution to higher-spin charges.

The latter depend on the field components and on the parameters of large gauge transformations preserving the leading falloffs \eqref{boundary-cond-s} as \cite{charges_spin-s,Campoleoni:2016uwr}
\be \label{charges-s} 
\begin{split}
Q(u) & = -\,\lim_{r\to\infty} \int_{S_u} \frac{r^{n-1} d\Omega_n}{(s-1)!}\, \sum_{p\,=\,0}^{s-1} \bin{s-1}{p}\, \bigg\{\, \vf_{u_{s-p}i_{p}}\left( r\pr_r + n + 2p \right) \epsilon^{u_{s-p-1}i_p} \\
& +  \epsilon^{u_{s-p-1}i_p} \bigg[\, (s-p-2)  \left( r\pr_r + n \right) \vf_{u_{s-p}i_{p}} - \frac{s-p-1}{r}\, D\cdot \vf_{u_{s-p-1}i_p} \,\bigg]\, \bigg\} \, .
\end{split}
\ee
The spin-$s$ asymptotic symmetries are rather involved and they have been fully exhibited in \cite{super,noi_charges} only for $s = 3$. Nevertheless, the charges \eqref{charges-s} only depend on the leading order of the components $\epsilon_{r_{s-k-1}i_k}$ of the gauge parameter, that is given by
\be \label{K-u}
\epsilon_{r_{s-k-1}i_k} = r^{2k} \sum_{m\,=\,0}^{s-k-1} c_{s}[k,m]\, u^m (D\cdot)^m K_{i_k}{}^{\!\!\!(k+m)}(x^j) \, ,
\ee
where the tensors $K_{i_k}{}^{\!\!\!(k)}$ generalise the tensors $T$, $\r_i$ and $K_{ij}$ that we encountered in subsection~\ref{sec:superrotations}. In particular, they are bound to satisfy appropriate differential equations spelled out in \cite{noi_charges}.
Combining \eqref{coulomb-s} and \eqref{K-u} one can see that the contribution of the Coulombic branch to the charges is manifestly finite. The one of the radiation branch would instead naively diverge. On the other hand, one can check that for $D > 4$ the contribution of radiation actually vanishes on shell thanks to the interplay between the relations imposed by the equations of motion and the differential constraints on the gauge parameters. Moreover, in the same setup the asymptotic symmetries actually coincide with the solutions of the (a priori more restrictive) Killing tensor equation $\d \vf_{\m_s} = 0$.
When $D > 4$ the final expression for the spin-$s$ conserved charge therefore reads
\be \label{charges-final}
Q = \int_{S^n} d\Omega_n \sum_{q\,=\,0}^{s-1} d_{s}[q]\, K^{(q)}_{i_q} \cM^{(q)\,i_q} \, ,
\ee
where the $d_{s}[q]$ are some normalisation constants fixed by the conventions that we have chosen in \eqref{coulomb-s} and \eqref{K-u}. The $u-$dependence precisely cancels, as required by the conservation of the charges \eqref{charges-s}, that is guaranteed when the fields are on-shell and the asymptotic symmetries coincide with the Killing symmetries of the model under investigation. 

Let us notice that \eqref{charges-final} does not reproduce the $Q^+_s$ in \eqref{Q-total} when one sets all $K^{(q)}$ with $q>0$ to zero. The four-dimensional charge $Q^+_s$ indeed depends on the radiation branch, and this is crucial in the derivation of Weinberg's soft theorem. A contribution of the $u-$dependent data of radiation is allowed because in this case the asymptotic symmetries do not coincide with the solutions of the Killing tensor equation. In particular, the function $K^{(0)} \equiv T$ is not anymore constrained as for Killing tensors, and the corresponding missing cancellations allow for a dependence on the radiation branch. For more details we refer to section 5 of \cite{noi_charges}.


\begin{thebibliography}{}
%
%
%
%

\bibitem{super}  
A.~Campoleoni, D.~Francia and C.~Heissenberg,
``{\it On higher-spin supertranslations and superrotations},''
JHEP {\bf 1705} (2017) 120
[\href{https://arxiv.org/abs/1703.01351}{arXiv:1703.01351 [hep-th]}]. 

\bibitem{noi_charges}
A.~Campoleoni, D.~Francia and C.~Heissenberg,
``{\it Asymptotic Charges at Null Infinity in Any Dimension},''
Universe {\bf 4} (2018) no.3,  47
[\href{https://arxiv.org/abs/1712.09591}{arXiv:1712.09591 [hep-th]}].

\bibitem{fronsdal}
C.~Fronsdal,
``{\it Massless Fields with Integer Spin},''
Phys.\ Rev.\ D {\bf 18} (1978) 3624.

\bibitem{Weinberg_64} 
S.~Weinberg,
``{\it Photons and Gravitons in $S$-Matrix Theory: Derivation of Charge Conservation and Equality of Gravitational and Inertial Mass},''
Phys.\ Rev.\  {\bf 135} (1964) B1049.

\bibitem{Weinberg_65} 
S.~Weinberg,
``{\it Infrared photons and gravitons},''
Phys.\ Rev.\  {\bf 140} (1965) B516.

\bibitem{Strominger_Invariance} 
A.~Strominger,
``{\it On BMS Invariance of Gravitational Scattering},''
JHEP {\bf 1407} (2014) 152
[\href{http://arxiv.org/abs/arXiv:1312.2229}{arXiv:1312.2229 [hep-th]}].

\bibitem{Strominger_Weinberg} 
T.~He, V.~Lysov, P.~Mitra and A.~Strominger,
``{\it BMS supertranslations and Weinberg's soft graviton theorem},''
JHEP {\bf 1505} (2015) 151
[\href{http://arxiv.org/abs/arXiv:1401.7026}{arXiv:1401.7026 [hep-th]}].


\bibitem{BMS} 
H.~Bondi, M.~G.~J.~van der Burg and A.~W.~K.~Metzner,
``{\it Gravitational waves in general relativity. VII. Waves from axisymmetric isolated systems},''
Proc.\ Roy.\ Soc.\ Lond.\ A {\bf 269} (1962) 21.

\bibitem{Sachs_Waves} 
R.~K.~Sachs,
``{\it Gravitational waves in general relativity. VIII. Waves in asymptotically flat space-times},''
Proc.\ Roy.\ Soc.\ Lond.\ A {\bf 270} (1962) 103.

\bibitem{Sachs_Symmetries} 
R.~Sachs,
``{\it Asymptotic symmetries in gravitational theory},''
Phys.\ Rev.\  {\bf 128} (1962) 2851.

\bibitem{Geroch_Lectures} 
R. Geroch, ``{\it Asymptotic Structure of Space-Time},'' in: ``Asymptotic Structure of Space-Time'', F. P. Esposito and L. Witten (eds.), Plenum Press, New York, 1977, 1-105.

\bibitem{Ashtekar_Lectures} 
A. Ashtekar, ``{\it Asymptotic Quantization},'' based on 1984 Naples Lectures, Bibliopolis, edizioni di Filosofia e Scienze, Napoli, 1987.

\bibitem{Strominger_YM} 
A.~Strominger,
``{\it Asymptotic Symmetries of Yang-Mills Theory},''
JHEP {\bf 1407} (2014) 151
[\href{http://arxiv.org/abs/arXiv:1308.0589}{arXiv:1308.0589 [hep-th]}].

\bibitem{Strominger_QED} 
T.~He, P.~Mitra, A.~P.~Porfyriadis and A.~Strominger,
``{\it New Symmetries of Massless QED},''
JHEP {\bf 1410} (2014) 112
[\href{http://arxiv.org/abs/arXiv:1407.3789}{arXiv:1407.3789 [hep-th]}].

\bibitem{Campiglia_QED} 
M.~Campiglia and A.~Laddha,
``{\it Asymptotic symmetries of QED and Weinberg's soft photon theorem},''
JHEP {\bf 1507} (2015) 115
[\href{http://arxiv.org/abs/arXiv:1505.05346}{arXiv:1505.05346 [hep-th]}].

\bibitem{Avery-Schwab} 
S.~G.~Avery and B.~U.~W.~Schwab,
``{\it Noether's second theorem and Ward identities for gauge symmetries},''
JHEP {\bf 1602} (2016) 031
[\href{http://arxiv.org/abs/arXiv:1510.07038}{arXiv:1510.07038 [hep-th]}].


\bibitem{Barnich_Revisited} 
G.~Barnich and C.~Troessaert,
``{\it Symmetries of asymptotically flat 4 dimensional spacetimes at null infinity revisited},''
Phys.\ Rev.\ Lett.\  {\bf 105} (2010) 111103
[\href{http://arxiv.org/abs/arXiv:0909.2617}{arXiv:0909.2617 [gr-qc]}].

\bibitem{Barnich_BMS/CFT} 
G.~Barnich and C.~Troessaert,
``{\it Aspects of the BMS/CFT correspondence},''
JHEP {\bf 1005} (2010) 062
[\href{http://arxiv.org/abs/arXiv:1001.1541}{arXiv:1001.1541 [hep-th]}].

\bibitem{Barnich_Charge} 
G.~Barnich and C.~Troessaert,
``{\it BMS charge algebra},''
JHEP {\bf 1112} (2011) 105
[\href{http://arxiv.org/abs/arXiv:1106.0213}{arXiv:1106.0213 [hep-th]}].

\bibitem{Hamada:2018vrw}
  Y.~Hamada and G.~Shiu,
  ``{\it Infinite Set of Soft Theorems in Gauge-Gravity Theories as Ward-Takahashi Identities},''
  Phys.\ Rev.\ Lett.\  {\bf 120} (2018) no.20,  201601
  [\href{https://arxiv.org/abs/1801.05528}{arXiv:1801.05528 [hep-th]}].

\bibitem{Strominger_rev}
A.~Strominger,
``{\it Lectures on the Infrared Structure of Gravity and Gauge Theory},'' Princeton University Press, Princeton, 2018, 1-200
[\href{https://arxiv.org/abs/1703.05448}{arXiv:1703.05448 [hep-th]]}.

\bibitem{Einstein-YM Barnich}
G.~Barnich and P.~H.~Lambert,
``{\it Einstein-Yang-Mills theory: Asymptotic symmetries},''
Phys.\ Rev.\ D {\bf 88} (2013) 103006
[\href{https://arxiv.org/pdf/1310.2698.pdf}{arXiv:1310.2698 [hep-th]}].

\bibitem{Maxwell d=3 Barnich}
G.~Barnich, P.~H.~Lambert and P.~Mao,
``{\it Three-dimensional asymptotically flat Einstein-Maxwell theory},''
Class.\ Quant.\ Grav.\  {\bf 32} (2015) no.24,  245001
[\href{https://arxiv.org/pdf/1503.00856.pdf}{arXiv:1503.00856 [gr-qc]}].

\bibitem{Strominger "Kac"}
T.~He, P.~Mitra and A.~Strominger,
``{\it 2D Kac-Moody Symmetry of 4D Yang-Mills Theory},''
JHEP {\bf 1610} (2016) 137
[\href{https://arxiv.org/pdf/1503.02663.pdf}{arXiv:1503.02663 [hep-th]}].

\bibitem{Adamo Casali}
T.~Adamo and E.~Casali,
``{\it Perturbative gauge theory at null infinity},''
Phys.\ Rev.\ D {\bf 91} (2015) no.12,  125022
[\href{https://arxiv.org/pdf/1504.02304.pdf}{arXiv:1504.02304 [hep-th]}].

\bibitem{Mao_em}
P.~Mao, H.~Ouyang, J.~B.~Wu and X.~Wu,
``{\it New electromagnetic memories and soft photon theorems},''
Phys.\ Rev.\ D {\bf 95} (2017) no.12,  125011
[\href{https://arxiv.org/abs/1703.06588}{arXiv:1703.06588 [hep-th]}].

\bibitem{Mao_note}
P.~Mao and J.-B.~Wu,
``{\it Note on asymptotic symmetries and soft gluon theorems},''
Phys.\ Rev.\ {\bf D 96} (2017) 065023 
[\href{https://arxiv.org/abs/1704.05740}{arXiv:1704.05740 [hep-th]}].

\bibitem{Strominger Color}
M.~Pate, A.~M.~Raclariu and A.~Strominger,
  ``{\it Color Memory: A Yang-Mills Analog of Gravitational Wave Memory},''
  Phys.\ Rev.\ Lett.\  {\bf 119} (2017) no.26,  261602
  [\href{https://arxiv.org/pdf/1707.08016.pdf}{arXiv:1707.08016 [hep-th]}].

\bibitem{hollands2004}
S.~Hollands and R.~M.~Wald,
``{\it Conformal null infinity does not exist for radiating solutions in odd spacetime dimensions},''
Class.\ Quant.\ Grav.\  {\bf 21} (2004) 5139
[\href{https://arxiv.org/abs/gr-qc/0407014}{gr-qc/0407014.}]

\bibitem{gravity_anyD}
K.~Tanabe, S.~Kinoshita and T.~Shiromizu,
``{\it Asymptotic flatness at null infinity in arbitrary dimensions},''
Phys.\ Rev.\ D {\bf 84} (2011) 044055
[\href{https://arxiv.org/abs/1104.0303}{arXiv:1104.0303 [gr-qc]}].

\bibitem{gravity_evenD_2}
S.~Hollands, A.~Ishibashi and R.~M.~Wald,
``{\it BMS Supertranslations and Memory in Four and Higher Dimensions},''
Class.\ Quant.\ Grav.\  {\bf 34} (2017) no.15,  155005
[\href{https://arxiv.org/abs/1612.03290}{arXiv:1612.03290 [gr-qc]}].

\bibitem{angular-momentum}
K.~Tanabe, T.~Shiromizu and S.~Kinoshita,
``{\it Angular momentum at null infinity in higher dimensions},''
Phys.\ Rev.\ D {\bf 85} (2012) 124058
[\href{https://arxiv.org/abs/1203.0452}{arXiv:1203.0452 [gr-qc]]}.

\bibitem{Pate_Memory}
M.~Pate, A.~M.~Raclariu and A.~Strominger,
  ``{\it Gravitational Memory in Higher Dimensions},''
  JHEP {\bf 1806} (2018) 138
  [\href{https://arxiv.org/abs/1712.01204}{arXiv:1712.01204 [hep-th]}].

\bibitem{Kapec:2015vwa}
D.~Kapec, V.~Lysov, S.~Pasterski and A.~Strominger,
``{\it Higher-Dimensional Supertranslations and Weinberg's Soft Graviton Theorem},''
Annals of Mathematical Sciences and Applications, Volume 2 (2017),
pp 69-94
[\href{https://arxiv.org/abs/1502.07644}{arXiv:1502.07644 [gr-qc]}].

\bibitem{Garfinkle:2017fre}
D.~Garfinkle, S.~Hollands, A.~Ishibashi, A.~Tolish and R.~M.~Wald,
``{\it The Memory Effect for Particle Scattering in Even Spacetime Dimensions},''
Class.\ Quant.\ Grav.\  {\bf 34} (2017) no.14,  145015
[\href{https://arxiv.org/abs/1702.00095}{arXiv:1702.00095 [gr-qc]}].

\bibitem{Mao_evenD}
P.~Mao and H.~Ouyang,
``{\it Note on soft theorems and memories in even dimensions},''
Phys.\ Lett.\ B {\bf 774} (2017) 715
[\href{https://arxiv.org/abs/1707.07118}{arXiv:1707.07118 [hep-th]}].

\bibitem{Campiglia_scalars}
M.~Campiglia and L.~Coito,
  ``{\it Asymptotic charges from soft scalars in even dimensions},''
  Phys.\ Rev.\ D {\bf 97} (2018) no.6,  066009
  [\href{https://arxiv.org/abs/1711.05773}{arXiv:1711.05773 [hep-th]}].

\bibitem{ACDmemory}
A.~Campoleoni, D.~Francia and C.~Heissenberg -- 
In preparation

\bibitem{Sagnotti_review}
A.~Sagnotti,
``{\it Notes on Strings and Higher Spins},''
J.\ Phys.\ A {\bf 46} (2013) 214006
[\href{http://arxiv.org/abs/arXiv:1112.4285}{arXiv:1112.4285 [hep-th]}].

\bibitem{lowspin}
D.~Francia,
``{\it Low-spin models for higher-spin Lagrangians},''
Prog.\ Theor.\ Phys.\ Suppl.\  {\bf 188} (2011) 94
[\href{https://arxiv.org/abs/1103.0683}{arXiv:1103.0683 [hep-th]}].

\bibitem{ML}
A.~Campoleoni and D.~Francia,
``{\it Maxwell-like Lagrangians for higher spins},''
JHEP {\bf 1303} (2013) 168
[\href{https://arxiv.org/abs/1206.5877}{arXiv:1206.5877 [hep-th]}].

\bibitem{carlo_tesi} 
C.~Heissenberg,
``{\it Asymptotic symmetries of gravity and higher-spin theories},''
Master Thesis, Scuola Normale Superiore and Universit\`a di Pisa, 2016. \href{https://etd.adm.unipi.it/t/etd-08172016-184514/}{ETD}

\bibitem{conformal-killing}
M.~G.~Eastwood,
``{\it Higher symmetries of the Laplacian},''
Annals Math.\  {\bf 161} (2005) 1645
[\href{http://arxiv.org/abs/hep-th/0206233}{hep-th/0206233}].

\bibitem{flat-algebras_1}
X.~Bekaert,
``{\it Comments on higher-spin symmetries},''
Int.\ J.\ Geom.\ Meth.\ Mod.\ Phys.\  {\bf 6} (2009) 285
[\href{http://arxiv.org/abs/arXiv:0807.4223}{arXiv:0807.4223 [hep-th]}].

\bibitem{flat-algebras_2}
C.~Sleight and M.~Taronna,
``{\it Higher-Spin Algebras, Holography and Flat Space},''
JHEP {\bf 1702} (2017) 095
[\href{http://arxiv.org/abs/arXiv:1609.00991}{arXiv:1609.00991 [hep-th]}].

\bibitem{charges_spin-s}
G.~Barnich, N.~Bouatta and M.~Grigoriev,
``{\it Surface charges and dynamical Killing tensors for higher spin gauge fields in constant curvature spaces},''
JHEP {\bf 0510} (2005) 010
[\href{http://arxiv.org/abs/hep-th/0507138}{hep-th/0507138}].

\bibitem{Campoleoni:2016uwr}
  A.~Campoleoni, M.~Henneaux, S.~H{\"o}rtner and A.~Leonard,
  ``{\it Higher-spin charges in Hamiltonian form. I. Bose fields},''
  JHEP {\bf 1610} (2016) 146
  [\href{http://arxiv.org/abs/arXiv:1608.04663}{arXiv:1608.04663 [hep-th]}].

\end{thebibliography}
%

\end{fmffile}
\end{document}